# Bitcoin Risk Modeling with Blockchain Graphs

Cuneyt Gurcan Akcora,[*] Matthew F. Dixon,[†] Yulia R. Gel[‡] and Murat Kantarcioglu[§]

## Abstract

A key challenge for Bitcoin cryptocurrency holders, such as startups using ICOs to raise funding, is managing their FX risk. Specifically, a mis-informed decision to convert Bitcoin to fiat currency could, by itself, cost USD millions.

In contrast to financial exchanges, Blockchain based crypto-currencies expose the entire transaction history to the public. By processing all transactions, we model the network with a high fidelity graph so that it is possible to characterize how the flow of information in the network evolves over time. We demonstrate how this data representation permits a new form of microstructure modeling — with the emphasis on the local topological network structure to study the role of users, entities and their interactions in formation and dynamics of crypto-currency investment risk. In particular, we identify certain sub-graphs (chainlets) that exhibit predictive influence on Bitcoin price and volatility and characterize the types of chainlets that signify extreme losses.

## 1 Introduction

Nascent empirical research suggests that short-run Bitcoin price behavior is prone to bubbles and busts and somewhat detached from asset pricing theory [9, 10, 5]. Since Bitcoin derives its economic value from a "network effect" – the more individuals who use Bitcoin, the more valuable the entire Bitcoin ecosystem becomes- it is expected that transaction activity is strongly linked with Bitcoin price changes [9, 10]. In contrast to existing financial networks, Blockchain based crypto-currencies expose the entire transaction graph to the public. Bitcoin transactions are listed for all participants and the most significant agents can be immediately located on the network.

In contrast to closed financial systems, the largest accounts in a crypto-currency exchange are listed and can be tracked over time and have been popularly referred to as "whales". The econometrics of Bitcoin seeks new interdisciplinary research to demonstrate how full disclosure of an agent's actions in a crypto-currency market inform price discovery and ultimately serve as an early-warning indicator for excess market volatility or even a crash.

With the goal of building a predictive model, we therefore depart from classical times-series cross sectional models that leverage standard macro economic variables such as GDP and inflation. Instead, we use Bitcoin's microstructure.

By processing all financial interactions, our objective is to model the network with a high fidelity graph so that it is possible to characterize how the flow of information in the network evolves over time. This novel data representation permits an entirely new form of financial econometrics — with the emphasis on the topological network structures rather than covariance of historical time series of prices. The role of users, entities and their interactions in formation and dynamics of crypto-currency risk investment, financial predictive analytics and, more generally, in re-shaping the modern financial world is a novel area of research [11, 6, 7, 8, 12].

## 2 Method: Chainlets and Data Processing

As shown in Figure 1, a Bitcoin graph consists of three main components: addresses, transactions and blocks (see [1] for a primer on Blockchain graphs). One approach to understand how transactions relate to market price is to introduce the novel concept of *k-chainlets* [2].

A $k$-chainlet is a Bitcoin sub-graph of $k \geq 1$ transactions and their corresponding input and output addresses corresponding to different accounts, not necessarily unique to a user. In the simplest case, a single transaction creates a 1-chainlet with one or more inputs and a single output. For example, in Figure 1, transaction $t_2$ results in the transfer of Bitcoin from addresses $a_3$, $a_4$, $a_5$ to address $a_8$. Such a transaction creates a 1-chainlet that has three inputs and one output. We denote this subgraph as a chainlet of type $\mathbb{C}_{3 \to 1}$, where 3 and 1 are the number of input and output addresses, respectively.

A 1-chainlet is the smallest building block of the Bitcoin graph; inputs and outputs of the chainlet are determined at once, and the transaction is digitally signed. This signed information cannot be modified, but multiple 1-chainlets can be combined to extend the graph. For simplicity, in the rest of this work, we use the term *chainlet* to refer to 1-chainlets.

[*]Postdoctoral Fellow, Data Security and Privacy Lab, University of Texas at Dallas.
[†]Assistant Professor of Finance & Statistics, Stuart School of Business, Illinois Institute of Technology.
[‡]Professor of Mathematics, Department of Mathematical Sciences, University of Texas at Dallas.
[§]Professor of Computer Science, Director of the Data Security and Privacy Lab, University of Texas at Dallas.

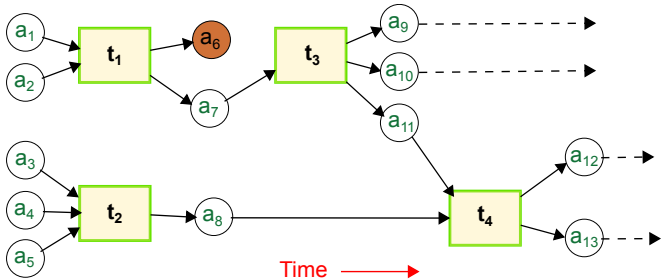

Figure 1: *A transaction-address graph representation of the Bitcoin network. Addresses are represented by circles, transactions with rectangles and edges indicate a transfer of coins. Blocks order transactions in time, whereas each transaction with its input and output nodes represents an immutable decision that is encoded as a subgraph on the Bitcoin network. Some addresses, such as $a_6$ in the figure, may contain unspent bitcoins.*

Graph analysis allows us to evaluate the local topological structure of the Bitcoin graph over time and assess the role of chainlets on Bitcoin price formation and dynamics.

Figure 2 illustrates how the activity of the network can be represented by a chainlet matrix. On a given day, we count the occurrences of each $\mathbb{C}_{i \to j}$ and store it in a chainlet matrix. The maximum number of inputs or outputs of a chainlet can be large, however, sometimes exceeding 1000. When the number of inputs and/or outputs exceeds a threshold $N$, we refer to these chainlets as "extreme chainlets". In our historical analysis of daily snapshots, we choose $N = 20$, which corresponds to the 97.5 percentile of all chainlet occurrences.

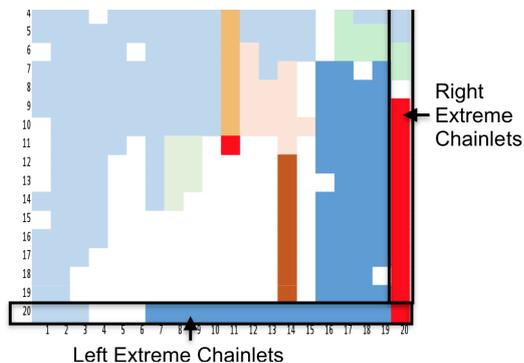

Figure 2: *This figure illustrates how the network is represented in time with a $20 \times 20$ chainlet matrix. Each matrix is formed by taking snapshots of the Bitcoin graph and counting the occurrences of $\mathbb{C}_{i \to j}$, $\forall i, j$ on a given day. The color scale denotes the frequency of chainlet occurrences. The left and right extreme chainlets are shown by the bottom row and far right column respectively.*

It is instructive to distinguish between 'left extreme chainlets' and 'right extreme chainlets'. Left extreme chainlets are the subset $\mathcal{C}^l := \{\mathbb{C}_{i \to j} \mid i = N, j \in \{1, \ldots, N\}\}$, as highlighted in the bottom row in Figure 2. They represent transactions from of a large number of accounts to fewer addresses. As a general rule, left extreme chainlets indicate bitcoin investment.

Right extreme chainlets are the subset $\mathcal{C}^r := \{\mathbb{C}_{i \to j} \mid i \in \{1, \ldots, N-1\}, j = N\}$, as highlighted in the far right column in Figure 2. They represent the sale of a large sum of Bitcoins across the market - the seller divides the balance and sends them to potentially hundreds of Bitcoin addresses. We denote the USD amount of Satoshis transferred on date $t$ by left and right extreme chainlets as $A_t^l$ and $A_t^r$, and the total occurrences as $O_t^l$ and $O_t^r$ respectively.

Figure 3 shows the ratio of extreme chainlets to total occurrences at time $t$, denoted as $O_t^x$. We also measure extreme chainlet activity with the ratio of Bitcoins transacted by extreme chainlets, $A_t^x$. For example, if the volume today was 2M Satoshis and 200K Satoshis were transacted by using extreme chainlets, then $A_t^x = 0.1$.

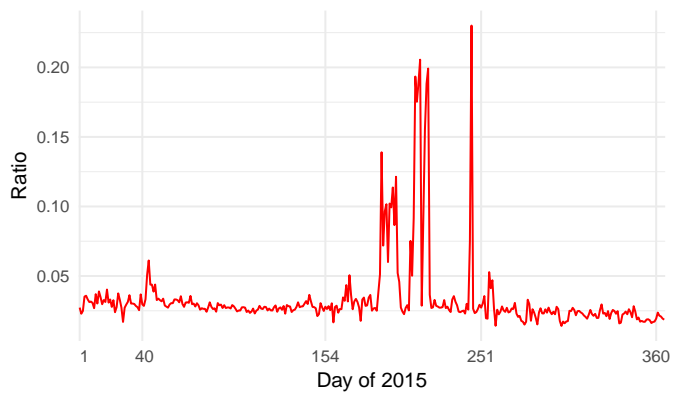

Figure 3: *Ratio of extreme chainlets by daily occurrence over 2015. On June 3 (day 154), New York State financial services superintendent announced BitLicense: a set of rules that would govern virtual-currency businesses. BitLicense came into effect on September 8th (day 251). Rather than complying with these rules, cryptocurrency exchanges demanded their customers to leave their platforms. Many customers left by selling their Bitcoins, as evidenced by high extreme chainlet activity.*

## 3 Forecasting Bitcoin

The extent to which we can build predictive models from the chainlets has already led to some promising results [2] (see [2] for specification of the types and groups of chainlets that exhibit predictive influence on Bitcoin price and volatility).

### 3.1 Risk modeling

We characterize the uncertainty of a 'loss' and, in particular, estimate the probability of extreme losses occurring over a future horizon. The loss is defined as the negative of the log returns, $L_t = -r_t$, where $r_t := ln(P_{t+1}/P_t)$ and $P_t$ is the Bitcoin price on day $t$.

Bitcoin prices are sourced from Coinbase over the period 1/1/2012 to 10/7/2017 (2107 observations)[1] and the corresponding chainlet matrices are available through the website [2].

Table 1 shows the results from regressing the square of the log returns $r_t^2$, a proxy for volatility, against $x_t$ – the vector of daily extreme chainlet activity.

|             | Estimate | Std. Error | t-value | $\Pr(>|t|)$ |     |
|-------------|----------|------------|---------|-------------|-----|
| (Intercept) | 0.9995   | 0.0807     | 12.385  | < 2e-16     | *** |
| $A_t^l$     | 0.7248   | 0.1063     | 6.818   | 1.21e-11    | *** |
| $A_t^r$     | 0.2959   | 0.1278     | 2.316   | 0.02068     | *   |
| $A_t^x$     | -0.5348  | 0.1313     | -4.073  | 4.82e-05    | *** |
| $O_t^l$     | -0.5699  | 0.1074     | -5.304  | 1.25e-07    | *** |
| $O_t^r$     | -0.4541  | 0.1644     | -2.762  | 0.00579     | **  |
| $O_t^x$     | 0.5043   | 0.1832     | 2.753   | 0.00595     | **  |

Table 1: *This table shows the statistical significance of the extreme chainlet regressors on $r_t^2$, a proxy for volatility. Note that both the response and the regressors have been standardized.*

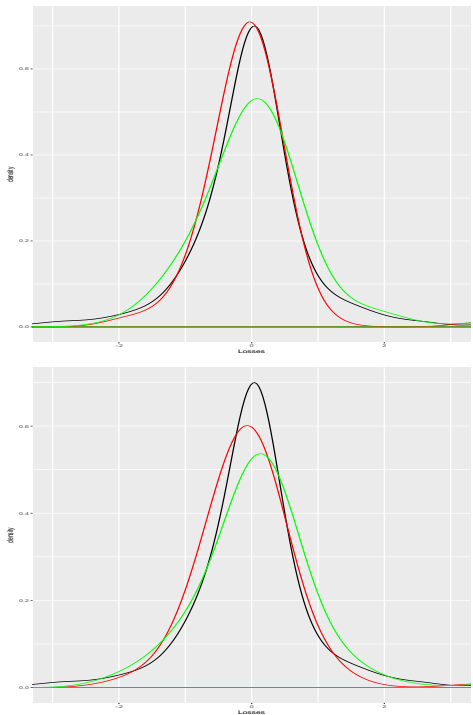

Figure 4: *The empirical densities of the standardized daily losses conditioned on the lower (red) and upper (green) $\alpha = 0.05$ percentiles of extreme chainlet activity by (top) amount ($A^x$) and (bottom) occurrences ($O^x$). The standardized unconditional loss density is shown by the black line.*

Results establishing Granger causality between prices and chainlets are shown in [2]. We emphasize that the purpose of our analysis here is to augment these results with the distributional properties of losses given extreme chainlet activity. Figure 4 and Table 2 show the unconditional loss densities, $\phi(L_t)$ (black) and conditional densities of the standardized daily Bitcoin losses over the same period from 1-1-2012 to 10-7-2017. The mean of the loss density is observed to shift to the right (indicating higher losses) when conditioned on the top fifth percentile (green) of extreme chainlet activity measures $A_t^x$ (top) and $O_t^x$ (bottom). Conversely, we observe that the mean of the loss density shifts to the left when conditioned on the lower fifth percentile (red) of extreme chainlet activity measures $A_t^x$ (top) and $O_t^x$ (bottom). The skew and kurtosis of the conditional loss distributions are also observed to differ significantly from the unconditional loss distribution.

[1] Our analysis (see Figure 3 in [2]) showed that the Bitcoin network did not stabilize until late 2011.
[2] https://github.com/cakcora/CoinWorks

| pdf | mean | std.dev. | skewness | kurtosis |
|---|---|---|---|---|
| $\phi(L_t)$ | 0 | 1 | 0.518 | 12.082 |
| $\phi(L_t \| A_t^x < \Phi_{A_t^x}^{-1}(0.05))$ | -0.047 | 1.107 | 3.283 | 31.618 |
| $\phi(L_t \| A_t^x > \Phi_{A_t^x}^{-1}(0.95))$ | 0.0861 | 0.843 | 1.590 | 6.046 |
| $\phi(L_t \| O_t^x < \Phi_{O_t^x}^{-1}(0.05))$ | -0.081 | 0.633 | 1.296 | 8.114 |
| $\phi(L_t \| O_t^x > \Phi_{O_t^x}^{-1}(0.95))$ | 0.118 | 0.930 | 2.025 | 10.457 |

Table 2: *This table show the moments of the conditional and unconditional empirical density functions corresponding to those shown in Figure 4.*

### 3.2 GARCH

The application of GARCH models to forecast Bitcoin has been extensively investigated in [4, 3]. We supplement these findings, by demonstrating the importance of including extreme chainlet activity, $x_t$ in the GARCH model. We choose an $ARMA(p,q) - GARCHX(1,1)$ model:

$$y_t = \mu + \sum_{i=1}^{p} y_{t-i} + \sum_{i=1}^{q} u_{t-1} + \sigma_t u_t, \quad (1)$$

$$\sigma_t^2 = \alpha_0 + \alpha_1 u_{t-1}^2 + \beta \sigma_{t-1}^2 + \beta_x^T x_t, \quad (2)$$

where $y_t$ are the observed daily returns, $u_t$ are standard skewed Student's $t$-innovations and $\sigma_t$ is the volatility. Additional diagnostics, provided on request, show a positive ARCH effect and that a GARCH(1,1) model has the lowest AIC with a ARMA(2,2) model for the mean equation. Both models pass a Box-Ljung and Lagrange Multiplier test at the 99% confidence level for the residuals and square of the residuals. Separate sign-bias tests show that asymmetry is not significant.

Table 3 compares the 99% daily Value-at-Risk (VaR) backtest of the ARMA(2,2)-GARCHX(1,1) model with the

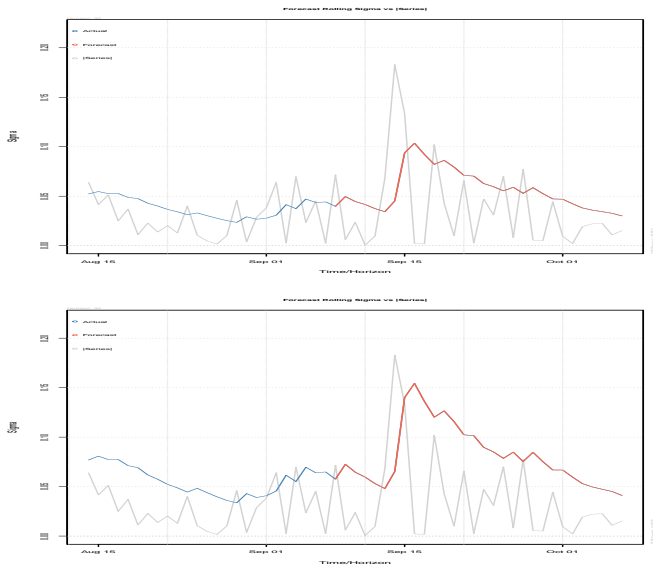

Figure 5: *Next day volatilities estimated using a (top) ARMA(2,2)-GARCH(1,1) model and a (bottom) ARMA(2,2)-GARCHX(1,1) model. The forecasting horizon is rolled over a 30 day out-of-sample period (red). In sample volatility estimates are shown in blue and observed daily returns are shown in gray.*

ARMA(2,2)-GARCH(1,1) model over an historical period of 1857 days. The backtest is performed over an in-sample rolling horizon, with model refitting every 7 days. The ARMA(2,2)-GARCHX(1,1) model is observed to pass the Kupiec unconditional and the Christoffersen conditional coverage tests at the 95% confidence level whereas the ARMA(2,2)-GARCH(1,1) underestimates the VAR and fails both tests.

|  | ARMA(2,2)-GARCH(1,1) | ARMA(2,2)-GARCHX(1,1) |
|---|---|---|
| alpha | 1% | |
| Expected Breaches | 18.6 | |
| Actual VaR Breaches | 33 | 15 |
| Unconditional Coverage (Kupiec) | | |
| $H_0$: Correct Breaches | | |
| LR.uc Statistic | 9.201 | 0.742 |
| LR.uc Critical | 3.841 | |
| LR.uc p-value | 0.002 | 0.389 |
| Reject Null | YES | NO |
| Conditional Coverage (Christoffersen) | | |
| $H_0$: Correct Breaches and Independence of Failures | | |
| LR.cc Statistic | 9.451 | 0.986 |
| LR.cc Critical | 5.991 | |
| LR.cc p-value | 0.009 | 0.611 |
| Reject Null | YES | NO |

Table 3: *This table compares coverage test results of VaR backtests using the GARCH and GARCHX model.*

Figure 5 compares the day ahead forecasted volatility, over a 30 day out-of-sample rolling horizon, (top) without and (bottom) with the chainlet regressors. The GARCHX model is observed to predict higher volatility than the GARCH model and is found to be preferable for modeling risk. Under a quadratic loss function, we reject $H_0$ in the Diebold-Mariano test (DM=-2.2728) and conclude that the differences in the model residuals are significant at the 95% level ($p = 0.023$).

## 4 Summary

We model the blockchain transaction history of Bitcoin with high fidelity graphs. Extreme chainlet activity, characterized by transaction amounts and occurrences are shown empirically to result in increased probability of losses and to significant changes in the volatility. With the inclusion of these chainlet activities as external regressors in the variance equation, we show a significant improvement in the GARCH model for predicting extreme next day losses. Orders in the mean and variance equation being equal, the inclusion of extreme chainlet regressors results in nearly 90% reduction in the number of false daily 99% VaR breaches or under-breaches over an approximately 5 year backtesting horizon.

## 5 Acknowledgements

This material is based upon work supported by the National Science Foundation under Grant No. IIS 1633331.